\documentclass[12pt, a4paper]{article}
\usepackage{cmap}
\usepackage[T2a, T1]{fontenc}
\usepackage[cp1251]{inputenc}
\usepackage[russian]{babel}
\usepackage[dvips]{graphicx}
\begin{document}
\title{Количественные методики восстановления истинных топографических свойств объектов по измеренным АСМ-изображениям:\protect\\
Часть 1. Контактные деформации зонда и образца}
\author{М.\,О.\,Галлямов, И.\,В.\,Яминский}
\date{\emph{
Физический факультет Московского государственного университета
им.\,М.\,В.\,Ломоносова, Москва}}
\maketitle
\begin{abstract}
Сделано предположение, что эффект занижения высот объектов исследования АСМ обусловлен контактными деформациями. Для количественного описания эффекта применено решение контактной задачи Герца. Построено общее численное решение задачи, а также, для случая цилиндрического образца, найдены приближенные аналитические решения, справедливые при определенных соотношениях параметров геометрии контакта. Обнаружено, что найденные решения хорошо согласуются с экспериментальными результатами. Разработанный подход позволил определить упругие параметры отдельного микрообъекта, адсорбированного на поверхность твердой подложки.
\end{abstract}

Несмотря на возможность достижения высокого пространственного разрешения, информация, получаемая методами зондовой микроскопии (в частности~--- атомно-силовой микроскопии (АСМ)), в ряде случаев неадекватно отображает реальные особенности поверхности вследствие \emph{артефактов} метода, обусловленных влиянием инструмента исследования на изучаемый \emph{объект}. Эти артефакты, как правило, легко учитываются на качественном уровне при интерпретации АСМ-результатов, однако специфика ряда задач может потребовать количественных оценок и методов восстановления реальной геометрии объектов.

Мы проанализировали два основных артефакта АСМ, влияние которых особенно существенно при проведении исследований \emph{отдельных} микрообъектов, адсорбированных на поверхность твердой подложки: эффекта уширения профиля (основные принципы методики учета этого эффекта частично изложены в работе\,\cite{1}) и эффекта занижения высот АСМ-изображений объектов исследования. Рассмотрение данного эффекта с позиций анализа контактных деформаций зонда и образца излагается ниже.

\section*{Контактные деформации}
С первых работ по АСМ-визуализации молекул нуклеиновых кислот\,\cite{2} отмечалось, что высоты АСМ-изображений ДНК существенно занижены в сравнении с имеющимися модельными представлениями о структуре молекулы. В то же время для ряда других объектов (с близкими физическими свойствами, но отличными радиусами кривизны) эффект занижения высот проявлялся не столь выражение. Так, в работах\,\cite{3, 4} были визуализованы вирусные частицы табачной мозаики (ВТМ) и молекулы вирусной РНК, причем было обнаружено, что эффект занижения высот для вирусных частиц несущественен, в то время, как высота АСМ-изображений молекул РНК занижена более чем на 50\%, несмотря на то, что и те и другие объекты были, как правило, визуализованы на одном кадре при одной силе сканирования. Применение излагаемой ниже методики позволило нам количественно описать данный эффект и связать его с различием радиусов частиц ВТМ ($\sim 10\,\mbox{нм}$) и нуклеиновых кислот ($< 1\,\mbox{нм}$).

Следуя\,\cite{5} мы предположили, что эффект занижения высот АСМ-изображений объектов связан с \emph{контактными деформациями}. Действительно, в процессе сканирования зонд и образец взаимодействуют с силами порядка $(1\div 100)\times 10^{-9}\,\mbox{Н}$ и, в силу малого радиуса кривизны зондирующего острия ($\sim 10\,\mbox{нм}$), оказывается, что контактное давление может составлять значительную величину и приводить к контактным деформациям.

\subsection*{Контакт двух тел}
Впервые задача о контактных деформациях двух тел была решена Г.\,Герцем в 1882\,г.\,\cite{6}, будем исходить из этого решения, изложенного, например, в\,\cite{7}. Если два контактирующих тела сдавливаются некоторой силой $F$, то они будут деформироваться и сблизятся на некоторое расстояние $h$, при этом областью соприкосновения уже будет не одна точка, а некоторый участок конечной площади $S$.

Анализ задачи включает рассмотрение суммарного тензора кривизны контактирующих поверхностей $\chi_{\alpha \beta}+\chi'_{\alpha \beta}$ главные значения которого $A$ и $B$ могут быть выражены через главные радиусы кривизны контактирующих поверхностей, соответствующие формулы для общего случая приведены в\,\cite{7}.

Решение \emph{контактной задачи}, при условии малости деформаций в сравнении с соответствующими радиусами кривизны, показывает, что формой области контакта является эллипс с полуосями $a$ и $b$, и позволяет выразить эти величины, а также сближение за счет деформации $h$, через известные параметры задачи: величину сдавливающей силы $F$, параметры геометрии контакта $A$ и $B$, а также коэффициент $D$, обратный эффективному модулю упругости:
\begin{equation}
\label{1}
D=\frac{3}{4}\left( \frac{1-\sigma^2}{E} + \frac{1-\sigma'^2}{E'} \right),
\end{equation}
здесь $E$, $E'$, $\sigma$ и $\sigma'$ модули Юнга и Пуассона материалов зонда и образца.

Однако в силу того, что конечные формулы решения \emph{контактной задачи} являются системой нелинейных уравнений с неявными зависимостями от искомых параметров $a$ и $b$ (см.\,\cite{7}), то для удобства применения этих соотношений при интерпретации экспериментальных результатов необходимы либо реализация численного решения, либо дополнительный анализ с привлечением упрощающих предпосылок. Ниже рассмотрим применение решения Герца к анализу важных для прикладных задач частных случаев.

\subsection*{Контакт сферического зонда и сферического (или плоского) образца}
Анализируемая здесь задача актуальна при рассмотрении контактных деформаций, возникающих при сканировании микрообъектов, форма которых может быть аппроксимирована сферой\footnote{например, молекул ряда белков и пр.}, а также плоских образцов, например, тонких пленок. 

Если зонд и образец вблизи точки контакта описываются сферическими поверхностями и характеризуются радиусами кривизны $R$ и $R'$, тогда,
$$
A=B=\frac{1}{2} \left( \frac{1}{R}+\frac{1}{R'} \right)
$$
откуда следует, что $a = b$ и соотношения, связывающие параметры задачи существенно упрощаются: легко показать, что область контакта будет представлять собой окружность радиуса $a$:
\begin{equation}
\label{2}
a=(FD)^{1/3}\left( \frac{1}{R} + \frac{1}{R'} \right)^{-1/3},
\end{equation}
здесь $D$ также описывается\,(\ref{1}).

Для величины $h$~--- сближения зонда и образца за счет контактной деформации~--- в этом случае справедлива формула:
\begin{equation}
\label{3}
h=(FD)^{2/3} \left(\frac{1}{R} + \frac{1}{R'}\right)^{1/3}
\end{equation}
В этих формулах, как и ранее, $F$~--- сила, сдавливающая зонд и образец.

Формулы (\ref{2}, \ref{3}) используются, например, авторами работы\,\cite{5} при проведении показательных оценок, весьма важных для адекватной интерпретации результатов АСМ-исследований (особенно биополимеров, характеризующихся невысокими значениями модуля Юнга: $E \sim 10^8 \div10^{10}\,\mbox{Па}$).

Однако указанные формулы являются следствием решения контактной задачи для частного случая, и неприменимы, например, для анализа контактных деформаций зонда и \emph{цилиндрического} образца.

\subsection*{Контакт сферического зонда и цилиндрического образца}

Именно модель цилиндрического образца следует рассматривать при анализе деформаций (в АСМ-исследованиях) микрочастиц цилиндрической формы (вирусных частиц, различных линейных макромолекул и пр.). 

Для случая контакта сферического зонда радиуса $R$ и боковой поверхности цилиндра (образца) радиуса $R'$ параметры $A$ и $B$ выражаются следующим образом:
\begin{equation}
\label{4}
A=\frac{1}{2} \left( \frac{1}{R} + \frac{1}{R'}\right), \qquad B=\frac {1}{2R}.
\end{equation}

Однако в этом случае соотношения, являющиеся решением контактной задачи, напрямую не упрощаются. Реализация численного решения возможна, но, в силу сложности решаемой системы, требует проведения предварительных аналитических преобразований. Поэтому мы провели дополнительные упрощения исходных соотношений и получили аналитические формулы для двух частных случаев (близких и различающихся значений параметров $A$ и $B$), которые могут быть полезны для оценок при интерпретации экспериментальных результатов. Сравнение решений, полученных по найденным приближенным формулам, с общим численным решением показало хорошее совпадение (при выполнении соответствующих условий приближений).

\subsubsection*{Случай различающихся главных значений суммарного тензора кривизны контактирующих поверхностей}

В случае контакта зонда и боковой поверхности цилиндра, при условии, что радиус цилиндра меньше радиуса зонда, из формулы\,(\ref{4}) следует, что главные значения суммарного тензора кривизны поверхностей различаются: $A > B$. Исходя из общих формул решения контактной задачи можно показать, что в этом случае $a < b$. В случае, когда это различие составляет достаточную величину, мы можем упростить исходные нелинейные интегральные соотношения (см.\,\cite{7}), воспользовавшись ассимптотикой полного эллиптического интеграла, справедливой при условии $a^2 \ll b^2$, что, очевидно, не является жестким условием:
\begin{equation}
\label{5}
K(k)=\ln \left(\frac{4}{k'} \right) + \dots,
\end{equation}
где $k'=\sqrt{1-k^2}$.
Тогда для сближения за счет деформации $h$ получим:
\begin{equation}
\label{6}
h=\left( \frac{4}{\pi^2C} \right)^{1/3}(C+1)\times(FD)^{2/3}\times B^{1/3},
\end{equation}
что по структуре совпадает с формулой\,(\ref{3}) для сферического случая. Здесь безразмерный параметр $C$ зависит, вообще говоря, от отношения параметров эллипса $a$ и $b$:
\begin{equation}
\label{7}
C=\ln \left( \frac{4b}{a}\right)-1=\frac{Bb^2}{Aa^2}.
\end{equation}
Из уравнения~(\ref{7}) при известном отношении $B/A$ можно численно определить отношение $b/a$, и, соответственно, значение безразмерного параметра $C$. Численное решение показывает, что значение параметра $C$ для многих задач лежит в диапазоне от 1 до 3, так, в частности, при анализе контакта зонда ($R=10\,\mbox{нм}$) и молекулы нуклеиновой кислоты ($R' = 1\,\mbox{нм}$), с достаточной точностью можно воспользоваться соотношением $C\simeq2$.

Формулы для параметров эллиптической области контакта $a$ и $b$ несколько громоздки и мы их не приводим, но по своей структуре они совпадают с уравнением\,(\ref{2}). Таким образом все искомые параметры могут быть \emph{непосредственно} выражены через известные величины ($F$, $D$, $A$, $B$) и параметр $C$, который можно определить из соотношения~(\ref{7}) или воспользоваться оценкой.

\subsubsection*{Случай близких главных значений суммарного тензора кривизны контактирующих поверхностей}
Случай близких значений величин $A$ и $B$ реализуется, например, для задачи контакта сферического зонда и боковой поверхности цилиндра при условии, что радиус цилиндра много больше радиуса зонда. Тогда, в силу соотношений~(\ref{4}), действительно $A\sim B$, и, можно показать, что $a\sim b$. В этом случае ассимптотика~(\ref{5}) теряет применимость и следует воспользоваться другой ассимптотикой полного эллиптического интеграла\,\cite{8}:
$$
K(k)=\frac{\pi}{2}(1+m)[1+\dots],
$$
где $m=(1-k')/(1+k')$, а $k'=\sqrt{1-k^2}$.
И в этом случае для параметров области контакта $a$ и $b$ можно вывести зависимости, совпадающие по структуре с~(\ref{2}), но здесь мы их также не приводим. Для сближения зонда и образца за счет деформации получим:
\begin{equation}
\label{8}
h=(FD)^{2/3}\times \left (\frac{1}{4A}+\frac{1}{4B}\right)^{-1/3},
\end{equation}
что, как и выше, имеет структуру, сходную с уравнениями~(\ref{3}) и~(\ref{6}).

Т.о. и для случая близких значений $A$ и $B$ могут быть получены приближенные соотношения, позволяющие найти искомые величины непосредственно по известным параметрам задачи.

Выше мы рассмотрели контактные деформации в области соприкосновения зонда и образца. Однако общая деформация, определяющая занижение высоты АСМ-профиля, включает еще и вклад деформаций в области контакта образца и подложки (имеется в виду случай, когда сверху на образец давит зонд). Для этого случая, нужно лишь соответствующим образом переопределить параметры $A$ и $B$, рассмотрев геометрию контакта образца радиуса $R'$ (на который сверху давит зонд радиуса $R$)\footnote{образец следует рассматривать в этом случае как изогнутый цилиндр с радиусом изгиба поверхности, контактирующей с подложкой: $R+2R'$} и плоской подложки:
\begin{equation}
\label{9}
A=\frac{1}{2R'}, \qquad B=\frac{1}{2}\left(\frac{1}{R+2R'}\right).
\end{equation}
Анализ этого случая не отличается от проведенного выше для значений $A$ и $B$, определяемых формулой~(\ref{4}).

\section*{Применение разработанного алгоритма к сравнительному анализу деформаций образцов с различными значениями радиусов}

В качестве теста мы применили разработанный алгоритм для вычисления контактных деформаций в модельных случаях цилиндрического образца с радиусом $1\,\mbox{нм}$ и $10\,\mbox{нм}$. Результаты приведены в таблице~1, где для удобства сравнительного анализа используются одинаковые параметры задачи.

\begin{table}
\begin{tabular}{|c|c|c|c|c|c|}
\hline
$R'$ & Область контакта & $a$ и $b$ & $P$ & $h$ & $\varepsilon$\\
\hline
1\,нм & зонд/образец & 0{,}46 и 2{,}2 нм & $1{,}6\times 10^9\,\mbox{Па}$ & 0{,}36\,нм & 18\% \\
& образец/подложка & 0{,}47 и 2{,}4 нм & $1{,}5\times 10^9\,\mbox{Па}$ &  0{,}34\,нм & 17\% \\
\hline
\multicolumn{4}{|r|}{суммарная деформация:} & 0{,}7\,нм & 35\% \\
\hline
10\,нм & зонд/образец & 1{,}1 и 1{,}8\,нм & $0{,}8\times 10^9\,\mbox{Па}$ & 0{,}29\,нм & 1{,}4\% \\
& образец/подложка & 1{,}3 и 2{,}7\,нм & $0{,}4\times 10^9\,\mbox{Па}$ & 0{,}21\,нм & 1\% \\
\hline
\multicolumn{4}{|r|}{суммарная деформация:} & 0{,}5\,нм & 2{,}5\% \\
\hline
\end{tabular}
\caption{Сравнительный анализ контактных деформаций для модели цилиндрического образца при двух значениях радиуса: 1 и 10 нм. Используемые параметры задачи: модуль Юнга образца $E' = 10^{10}\,\mbox{Па}$, зонда $E = 10^{11}\,\mbox{Па}$, величина сжимающей силы $F = 5\times 10^{-9}\,\mbox{Н}$ и радиус кривизны кончика зонда $R = 10\,\mbox{нм}$\protect\\
В столбцах таблицы указаны: $R'$~--- радиус образца, $a$ и $b$~--- параметры области контакта, $P$~--- контактное давление, $h$~--- величина сближения за счет деформации, $\varepsilon$~--- относительная деформация ($h/2R'\times 100\%$)}
\end{table}

В таблице приведены результаты численных расчетов точного решения. Расчеты по приближенным методикам дают следующие различия с точным решением: для случая $R'= 1\,\mbox{нм}$ приближенное решение для случая различающихся $A$ и $B$ дает отличие в значениях $a$ и $b$ около 2\%, и в значении $h$~--- 0{,}5\%; для случая $R'=10\,\mbox{нм}$ приближенное решение для случая близких $A$ и $B$ дает отличие от точного решения для $a$ и $b$ около 10\%, для $h$ около 1\%.

Основной вывод из результатов таблицы тот, что, при прочих равных условиях, относительные деформации объектов с \emph{меньшим} радиусом кривизны существенно \emph{выше}. Т.о. мы объяснили упомянутый выше экспериментальный эффект, проявляющийся в том, что \emph{относительные} деформации молекул нуклеиновых кислот существенно превышают \emph{относительные} деформации частиц ВТМ.

\subsection*{Сравнение с экспериментальными данными}

С целью экспериментальной проверки закона ``две третьих'' (см. формулы~(\ref{6}) и~(\ref{8})):
\begin{equation}
\label{10}
h\sim (FD)^{2/3}\times f(R,R’),
\end{equation}
 мы исследовали деформации вирусных частиц табачной мозаики и молекул ДНК при различных значениях нагружающей силы сканирования.

Для вирусных частиц табачной мозаики наблюдалось хорошее совпадение эксперимента с теорией (с законом ``две третьих''\,(\ref{10})), см. рис.\,1. Экспериментальные погрешности определены как стандартные отклонения средних арифметических при статистической обработке значений, полученных из анализа нескольких АСМ-изображений для конкретного значения силы сканирования.

\begin{figure}
\begin{center}
\includegraphics*[width = 0.8 \textwidth]{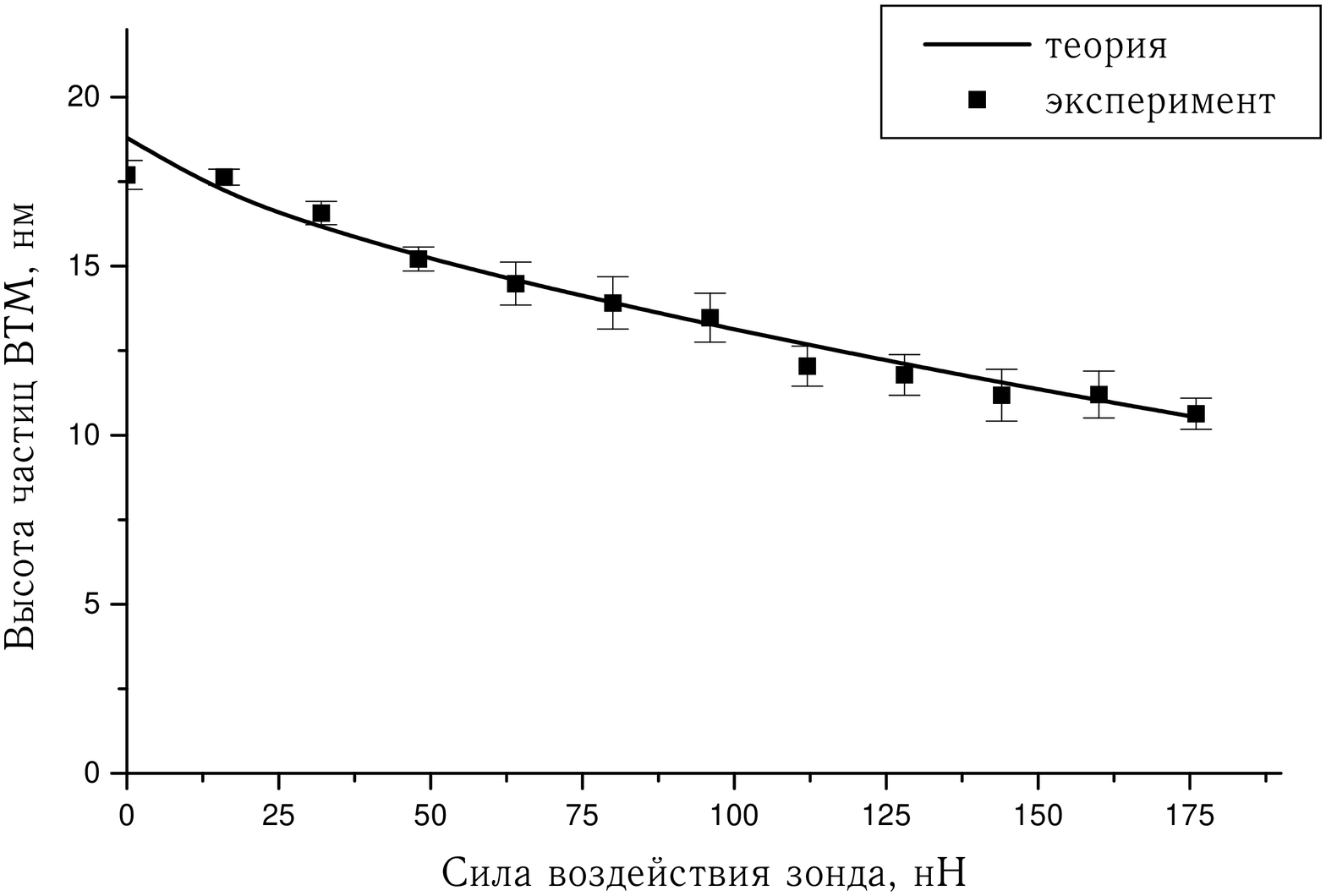}
\caption {Экспериментальная и теоретическая зависимости деформации частиц вируса табачной мозаики от величины нагружающей силы при сканировании.
}
\end{center}
\end{figure}

Теоретическая зависимость получена по рассмотренной методике анализа контактных деформаций цилиндрического образца (точное решение), где использовались значения $R = 25\,\mbox{нм}$ и $E' = 3\times 10^9\,\mbox{Па}$ (значения определены экспериментально). Из рис.\,1 следует, что закон ``две третьих''~(10) справедлив для исследуемого случая в широком диапазоне сил, за исключением области минимальных воздействий. Это может быть связано с тем, что в эксперименте на воздухе присутствие капиллярных сил не позволяет минимизировать силу воздействия зонда на образец до величины меньшей, чем несколько наноньютонов.

\begin{figure}
\begin{center}
\includegraphics*[width = 0.8 \textwidth]{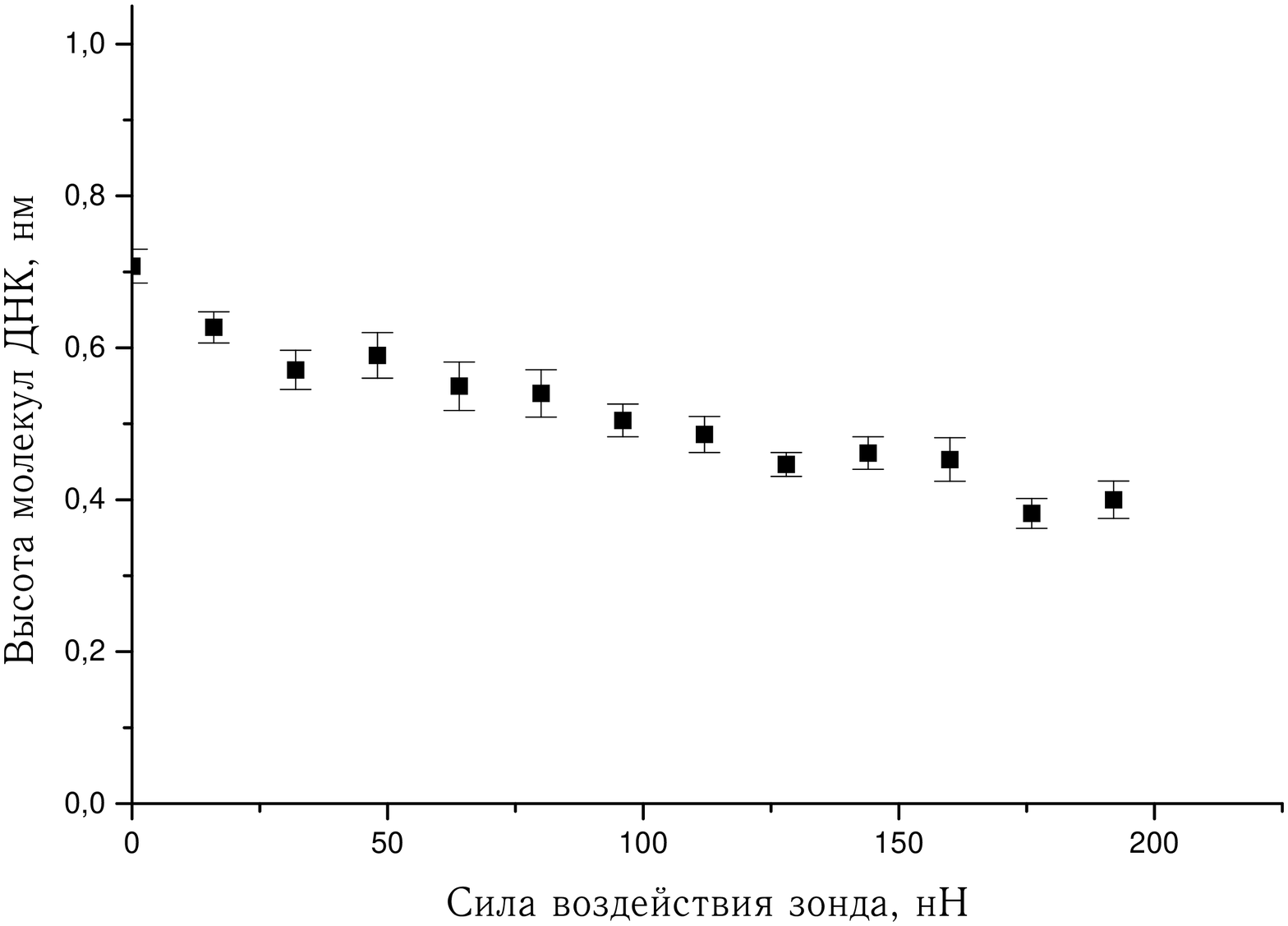}
\caption {Экспериментальная зависимость деформации молекул ДНК от величины нагружающей силы.
}
\end{center}
\end{figure}

Для молекул ДНК соответствующая экспериментальная зависимость изображена на рис.\,2. Эксперимент проводился по той же схеме и в тех же условиях, что и для случая анализа деформаций ВТМ. Однако в рассматриваемом случае деформации, по-видимому, не могут быть описаны законом ``две третьих''\,(\ref{10}). Вместо этого мы наблюдаем обычную линейную зависимость, т.е. закон Гука, что объясняется невыполнением в этом случае условия малости деформаций, при котором справедливы выводы контактной теории. То, что для случая молекул ДНК экспериментально измеренные относительные деформации велики даже при малых силах воздействия зонда, объясняется, опять же, присутствием капиллярных сил (капиллярного мостика), не позволяющих минимизировать силу сканирования на воздухе до значений меньших, чем несколько наноньютонов.

\emph{Благодарность.} Работа была поддержана РФФИ, проект \No\,97-03-32778a. Авторы выражают глубокую благодарность Ю.\,Ф.\,Дрыгину, за приготовление образцов ВТМ, а также В.\,В.\,Прохорову и Д.\,В.\,Клинову за любезно предоставленный образец ДНК.

\section*{Quantitative methods for deconvolution of true topographical properties of object on the basis of AFM-images:\protect\\
Part 1. Contact tip-sample deformations}
\subsection*{M.\,O.\,Gallyamov, I.\,V.\,Yaminsky}
We assume that the sample height measured using AFM is reduced due to contact deformation. The Herz contact theory is applied for the quantitative description. General numerical solution is found. Analytical approximations for specific contact geometry are derived for the case of cylindrical sample. It is found that theoretical description are consistent with experimental data. The developed approach has allowed to determine elastic parameters of individual microobjects adsorbed on a surface of solid substrate.


\begin{thebibliography}{99}
\bibitem{1} А.\,С.\,Андреева, М.\,О.\,Галлямов, О.\,А.\,Пышкина, В.\,Г.\,Сергеев, И.\,В.\,Яминский, // Журнал физической химии. 1999. Т.\,73. No.\,11. С.\,2062
\bibitem{2} С.\,Bustamante, J.\,Vesenka, С.\,L.\,Tang, W.\,Rees, M.\,Guthold, R.\,Keller // Biochemistry. 1992. V. 31. P. 22
\bibitem{3} Yu.\,F.\,Drygin, O.\,A.\,Bordunova, M.\,O.\,Gallyamov, I.\,V.\,Yaminsky // FEBS letters. 1998. V.425. P.217
\bibitem{4} M.\,О.\,Галлямов, Ю.\,Ф.\,Дрыгин, И.\,В.\,Яминский // Поверхность. 1999. No. 7. С.\, 104
\bibitem{5} Z.\,Shao, J.\,Mou, D.\,M.\,Czajkowsky, J.\,Yang, J.-Y.\,Yuan // Advances in Physics. 1996. V. 45. No. I.P.I
\bibitem{6} H.\,Herz // J. Reine Angew. Math. 1882. V. 92. Р. 156
\bibitem{7} Л.\,Д.\,Ландау, Е.\,M.\,Лифшиц. Теория упругости. M.: Наука, 1987. 246 с.
\bibitem{8} Г.\,Б.\,Двайт. Таблицы интегралов. M.: Наука, 1973. 228с.
\end{thebibliography}
\end{document}